\documentclass[A4paper,two column]{article}
\usepackage{graphicx}
\usepackage{amsmath}
\usepackage{amsfonts}

\newtheorem{lemma}{Lemma}
\begin{document}

\title{A hierarchy of entanglement criteria for four qubit symmetric Greenberger-Horne-Zeilinger diagonal states}
\author{Xiao-yu Chen \thanks{Email:xychen@zjgsu.edu.cn}, Li-zhen Jiang \\
{\small {College of Information and Electronic Engineering, Zhejiang Gongshang University, Hangzhou, 310018, Chna}}\\
}
\date{}
\maketitle

\begin{abstract}
  With a two step optimization method of entanglement witness, we analytically propose a set of necessary and sufficient entanglement criteria for four qubit symmetric Greenberger-Horne-Zeilinger (GHZ) diagonal states. The criterion set contains four criteria. Two of them are linear with density matrix elements. The other two criteria are nonlinear with density matrix elements. The criterion set has a nest structure. A proper subset of the criteria is necessary and sufficient for the entanglement of a proper subset of the states. We illustrate the nest structure of criterion set with the general Werner state set and its superset the highly symmetric GHZ diagonal state set, they are subsets of the symmetric GHZ diagonal state set.

PACS number(s): 03.65.Ud; 03.67.Mn;

\end{abstract}

\section{Introduction}
Entanglement plays a central role in quantum computation, quantum simulation and multipartite quantum communication. However, determining whether a given quantum state is entangled or not is by no means easy both in theoretical and experimental. Many criteria had been developed to detect entanglement\cite{Peres}\cite{Horodecki96}\cite{Rudolph}\cite{KChen}\cite{Guhne2004}\cite{Doherty}\cite{Li}\cite{Chen2018}, see Ref.\cite{GuhneToth} and \cite{Horodecki09} for overviews. A solution to the entanglement detection problem, known as entanglement witnessing, relies on the geometry of the set of all separable quantum states \cite{Horodecki96},\cite{Horodecki09}. The method of entanglement witness(EW) can easily be extended to multipartite cases \cite{Horodecki01}. Recent developments of the entanglement witness based criteria are entanglement witness for continuous variable system\cite{Gerke}, ultrafine entanglement witness\cite{Shahandeh1},entanglement witness game\cite{Zhou} and separability eigenvalue equation\cite{Sperling}. In principle, there exists the extremal EW\cite{Sperling} such that the entanglement criteria are necessary and sufficient. Practically, finding a solution to separability eigenvalue equation is still very difficult in multipartite scenario. Hence the aim of optimal detecting entanglement in a multipartite system is not easily reached in general. The proper starting point for such an aim is to investigate the states that are diagonal in GHZ basis\cite{Kay}.
GHZ diagonal states as special multipartite quantum states arise frequently in quantum information processing. GHZ diagonal states are tractable in many theoretical problems such as quantum channel capacity\cite{Chen2011}.
Most of multipartite entangled states prepared in experiments are GHZ states. Recently, there are experiments on four-qubit GHZ states: long-lived four qubit GHZ states are realized\cite{Kaufmann}, test of irreducible four qubit GHZ paradox has been produced\cite{Su}.
When imperfects in the preparation and decays are considered, the states prepared are usually GHZ diagonal states. The relationship of positive partial transpose (PPT) criterion and full separability of GHZ diagonal states had been studied \cite{Kay} and a simple condition had been given.
When the condition is not fulfilled, the border of full separability and entanglement may not be uncovered by PPT criterion. Then a complicated EW would be devised to detect the border. For three qubit GHZ diagonal states, the EW has been found \cite{GuhnePLA}\cite{Chen2015}\cite{Chen2017}
hence the necessary and sufficient criterion of full separability has been known. For the four qubit GHZ diagonal states, the full separability criterion had known for GHZ state mixed with white noise\cite{Dur}\cite{GuhneSeevinck} (also known as generalized Werner state \cite{Pittenger}). We may put the criterion in a criterion set $\mathcal{C}_{1}$. Let the set of all the generalized Werner states to be the state set $\mathcal{S}_{1}$. Then the criterion set $\mathcal{C}_{1}$ is necessary and sufficient for the full separability of the state set $\mathcal{S}_{1}$.

To detect the (multipartite) entanglement of multi-qubit systems (state set $\mathcal{S}_{N}$), we should consider a hierarchy of state sets.
The sets can be graph diagonal state set $\mathcal{S}_{5}$, GHZ diagonal state set $\mathcal{S}_{4}$, symmetric GHZ diagonal state set $\mathcal{S}_{3}$ , highly symmetric GHZ diagonal state set $\mathcal{S}_{2}$, generalized Werner state set $\mathcal{S}_{1}$. For four qubit system, we will build the criterion set $\mathcal{C}_{3}$ which is necessary and sufficient for state set $\mathcal{S}_{3}$ and it is a pretty good necessary criterion set for larger state set $\mathcal{S}_{j}$ with $j\geq4$. We will show that $\mathcal{C}_{1}\subset\mathcal{C}_{2}\subset\mathcal{C}_{3}$ for the state set inclusion relations $\mathcal{S}_{1}\subset\mathcal{S}_{2}\subset\mathcal{S}_{3}$.

We use a two steps procedure of finding the proper EW for a given four qubit GHZ state. The first step is to make the EW optimal in order to obtain necessary criterion of full separability, we will illustrate it in section II. The second step is to match the optimal EW with the state under investigation in order to obtain sufficient criterion of full separability, we will illustrate it in section III. Section IV and section V are devoted to state sets $\mathcal{S}_{2}$,$\mathcal{S}_{3}$ and their necessary and sufficient criterion sets $\mathcal{C}_{2},\mathcal{C}_{3}$, respectively. We discuss the relationship of PPT criterion and our criteria in section VI and conclude in section VII.

\section{Optimal entanglement witness}
Suppose there is a composed Hilbert space $\mathcal{H}=\mathcal{H}_{1}\otimes\cdot\cdot\cdot\otimes\mathcal{H}_{n}$. A quantum state $\sigma$ is called fully separable (hereafter abbreviated as `separable' sometimes), if it can be written as a classical mixture of product states\cite{Werner}:
\begin{equation}\label{1a}
\sigma=\sum_{i}q_{i}|\psi^{(i)}_{1}\rangle\langle\psi^{(i)}_{1}|\otimes\cdot\cdot\cdot\otimes|\psi^{(i)}_{n}\rangle\langle\psi^{(i)}_{n}|,
\end{equation}
with $q_{i}$ being a classical probability distribution, $|\psi^{(i)}_{j}\rangle$ is a pure state in the Hilbert space $\mathcal{H}_{j}$. A quantum state is entangled if it can be written as (\ref{1a}).

A four qubit GHZ diagonal state takes the form
\begin{equation}\label{1c1}
  \rho=\sum_{j=1}^{16}p_{j}|GHZ_{j}\rangle\langle GHZ_{j}|,
\end{equation}
where $p_{j}$ is a probability distribution. The GHZ state basis consists of sixteen vectors $|GHZ_{j}\rangle=\frac{1}{\sqrt{2}}(|0x_{2}x_{3}x_{4}\rangle \pm|1\overline{x}_{2}\overline{x}_{3}\overline{x}_{4}\rangle)$, with $x_{i}, \overline{x}_{i}\in\{0, 1\}$ and $x_{i} \neq \overline{x}_{i}$ . In
the binary notation, $j-1 = 0x_{2}x_{3}x_{4}$ for the `+' states and $j-1 = 1\overline{x}_{2}\overline{x}_{3}\overline{x}_{4}$ for the `-' states.

A four qubit symmetric GHZ diagonal state takes the form
   \begin{eqnarray}\label{1d1}
     \rho=p_{1}|GHZ_{1}\rangle\langle GHZ_{1}|+p_{2}\sum_{j=2,3,5,8}|GHZ_{j}\rangle\langle GHZ_{j}|\nonumber\\
     +p_{16}|GHZ_{16}\rangle\langle GHZ_{16}|
+p_{15}\sum_{j=9,12,14,15}|GHZ_{j}\rangle\langle GHZ_{j}|\nonumber\\
     +p_{4}\sum_{j=4,6,7}|GHZ_{j}\rangle\langle GHZ_{j}|+p_{13}\sum_{j=10,11,13}|GHZ_{j}\rangle\langle GHZ_{j}|\nonumber,
   \end{eqnarray}
with $p_{i}\geq0$ and normalization
\begin{equation}\label{1e}
     p_{1}+p_{16}+4(p_{2}+p_{15})+3(p_{4}+p_{13})=1.
\end{equation}

A four qubit highly symmetric GHZ diagonal state investigated in this paper takes the form
   \begin{eqnarray}\label{1d}
     \rho=p_{1}|GHZ_{1}\rangle\langle GHZ_{1}|+p_{16}|GHZ_{16}\rangle\langle GHZ_{16}|\nonumber\\
     +p_{2}\sum_{j=2}^{8}|GHZ_{j}\rangle\langle GHZ_{j}|+p_{15}\sum_{j=9}^{15}|GHZ_{j}\rangle\langle GHZ_{j}|
   \end{eqnarray}
is a special symmetric GHZ diagonal state with $p_{i}\geq0$ and normalization
\begin{equation}\label{1e}
     p_{1}+p_{16}+7(p_{2}+p_{15})=1.
\end{equation}

A generalized Werner state (A GHZ state mixed with white noise\cite{Pittenger} )
\begin{equation}\label{1f}
     \rho_{W}=p|GHZ\rangle\langle GHZ|+\frac{1-p}{16}\mathbf{I},
\end{equation}
is a special highly symmetric GHZ diagonal state, with $|GHZ\rangle=|GHZ_{1}\rangle$ and $\mathbf{I}$ being the $16\times 16$ identity matrix.

EW is a Hermite operator $\hat{W}$ such that $Tr\rho_{s}\hat{W}\geq 0$ for all separable state $\rho_{s}$ and $Tr\rho\hat{W}<0$ for at least one entangled state $\rho$. We may assume $\hat{W}=\Lambda\mathbb{I}-\hat{M}$,
where $\mathbb{I}$ is the identity operator and $\Lambda=\max_{\rho_s}Tr\rho_s\hat{M}$ such that
$\hat{W}$ is an optimal EW (the equality in $Tr\rho_{s}\hat{W}\geq 0$ can be reached). We may express the multi-qubit state and the EW with their characteristic functions.
Correspondingly, the operator $\hat{M}$ is characterized by real parameters $M_{i}$ $(i=1,...,4^{n}-1)$ in detecting the entanglement of a n-qubit state.
Here the number of parameters $M_{i}$ is equal to the number of free real parameters for describing the density matrix.
One of the widely used numeric method of finding a proper EW resorts to semi-definite programming.
The procedure of analytically finding a precise EW is divided into two steps. The first step is to find $\Lambda$ for the given $M_{i}$.
Notice that any operator $\hat{M}$ corresponds to a valid EW if $\Lambda$ is obtained. Hence the first step gives a valid necessary criterion of separability.
The second step is to adjust the parameters $M_{i}$ such that the EW detects all the entanglement. The parameters $M_{i}$ should match to the state under consideration,
so the second step gives the sufficient criterion of separability.

The two steps of finding entanglement criterion are just the two kinds of optimizations. The first step is the maximization to obtain $\Lambda$ (thus optimal EW)
for a given set of parameters $M_{i}$. The second step is to optimize with respect to $M_{i}$ such that the criterion is tight.

\begin{figure}[tpb]
\includegraphics[ trim=0.000000in 0.000000in -0.138042in 0.000000in,
height=2.5in, width=3.5in]{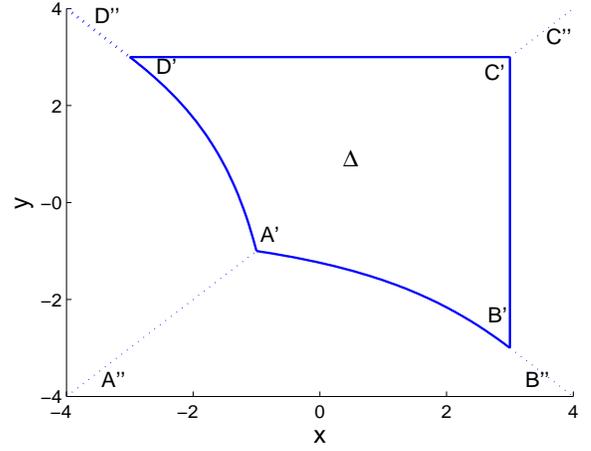}
\caption{The regions in $xy$-plane for the maximal mean of $\tilde{g}$ in $xy$-plane, here $x=\frac{M_{8}}{M_{9}},y=\frac{M_{15}}{M_{9}}$. $\Delta$ is bounded by two straight line sections and two curve sections. The two line sections are $x=3, y\in[-3,3]$ and $y=3, x\in[-3,3]$, respectively. The two curve sections are $y=3+\frac{1}{2}(\frac{9}{x}-x), x\in[-3,-1] $ and $x=3+\frac{1}{2}(\frac{9}{y}-y), y\in[-3,-1]$, respectively.}
\end{figure}
Let $\hat{M}$ be a Hermitian operator which is a linear combination of the tensor products of Pauli operators appearing in the four qubit GHZ diagonal states, namely,
\begin{eqnarray}\label{1}\nonumber
 \hat{M}&=&M_{1}IIZZ+M_{2}IZIZ+M_{3}IZZI+M_{4}ZIIZ\\
&&+M_{5}ZIZI+M_{6}ZZII\nonumber+M_{7}ZZZZ \\
&&+M_{8}XXXX+M_{9}XXYY\nonumber+M_{10}XYXY\\
&&+M_{11}XYYX+M_{12}YXXY+M_{13}YXYX\nonumber \\
&&+M_{14}YYXX+M_{15}YYYY.
\end{eqnarray}
Where $X,Y,Z$ are Pauli matrices, $I$ is the $2\times2$ identity matrix, $M_{i}$ are parameters mentioned above.

The mean of the operator $\hat{M}$ on
the pure product state $|\psi\rangle=|\psi_{1}\rangle|\psi_{2}\rangle|\psi_{3}\rangle|\psi_{4}\rangle$ is $\langle\psi|\hat{M}|\psi\rangle$.
Where $|\psi_{j}\rangle$ is a pure state of the $jth$ qubit, we parameterize it with angles $\theta_{j},\varphi_{j}$, namely, $|\psi_{j}\rangle=\cos\frac{\theta_{j}}{2}|0\rangle+\sin\frac{\theta_{j}}{2}e^{i\varphi_{j}}|1\rangle$.
 We may alternatively denote the mean as  $\langle\psi|\hat{M}|\psi\rangle =f(\boldsymbol{\theta},\boldsymbol{\varphi})$, where
 \begin{eqnarray}\label{2}\nonumber
   f(\boldsymbol{\theta},\boldsymbol{\varphi})=M_{1}z_{3}z_{4}+M_{2}z_{2}z_{4}+M_{3}z_{2}z_{3}+M_{4}z_{1}z_{4}\\
   M_{5}z_{1}z_{3}+M_{6}z_{1}z_{2}+M_{7}z_{1}z_{2}z_{3}z_{4}+g(\boldsymbol{\varphi})t_{1}t_{2}t_{3}t_{4}
 \end{eqnarray}
 with $z_{j}=\cos\theta_{j},t_{j}=\sin\theta_{j}$ and $\boldsymbol{\theta}=(\theta_{1},\theta_{2},\theta_{3},\theta_{4}),\boldsymbol{\varphi}=(\varphi_{1},\varphi_{2},\varphi_{3},\varphi_{4})$. The function $g(\boldsymbol{\varphi})$ is defined as
 \begin{eqnarray}\label{3}\nonumber
   g(\boldsymbol{\varphi})=M_{8}c_{1}c_{2}c_{3}c_{4}+M_{9}c_{1}c_{2}s_{3}s_{4}+M_{10}c_{1}s_{2}c_{3}s_{4}\\
   +M_{11}c_{1}s_{2}s_{3}c_{4}+M_{12}s_{1}c_{2}c_{3}s_{4}+M_{13}s_{1}c_{2}s_{3}c_{4}\nonumber\\
   +M_{14}s_{1}s_{2}c_{3}c_{4}+M_{15}s_{1}s_{2}s_{3}s_{4}.
 \end{eqnarray}
 with $c_{j}=\cos\varphi_{j},s_{j}=\sin\varphi_{j}$.
The maximization of the mean of operator $\hat{M}$ with respect to product states is transformed to the maximization of function $f(\boldsymbol{\theta},\boldsymbol{\varphi})$ with respect to the angles $\theta_{j}$,$\varphi_{j}$ (j=1,..,4). We can see that the maximization on $\varphi_{j}$ is independent of the maximization on $\theta_{j}$. The structure of the density matrix of GHZ diagonal states is 'X' type. The density matrix contains diagonal and anti-diagonal entries and all the other entries are zeros. In computational basis, the operator $\hat{M}$ can also be expressed as a matrix containing diagonal and anti-diagonal entries and all the other entries are zeros. The function $g(\boldsymbol{\varphi})$ is responsible for the property of anti-diagonal part of $\hat{M}$.

Denote $\tilde{g}=\max_{\boldsymbol{\varphi}}g(\boldsymbol{\varphi})$. Two of the angles can be removed by obvious optimization of triangle function. After some algebra, we have  $\tilde{g}=\max_{\varphi_{+},\varphi_{-}}g_{1}(\varphi_{+},\varphi_{-})$, with $\varphi_{\pm}=\varphi_{1}\pm\varphi_{2}$, and

   \begin{eqnarray}\label{4}\nonumber
     g_{1}(\varphi_{+},\varphi_{-})=\sqrt{(A_{1}c_{+}+A_{3}c_{-})^2+(A_{5}s_{+}-A_{7}s_{-})^2}\\
     +\sqrt{(A_{2}c_{+}+A_{4}c_{-})^2+(A_{6}s_{+}-A_{8}s_{-})^2}.
   \end{eqnarray}
   with $(A_{1},A_{2},A_{3},A_{4})=\frac{1}{2}(M_{8},-M_{9},-M_{14},M_{15})H$, $(A_{5},A_{6},A_{7},A_{8})=\frac{1}{2}(M_{10},M_{11},M_{12},M_{13})H$,
   $c_{\pm}=\cos\varphi_{\pm},s_{\pm}=\sin\varphi_{\pm}$, where $H$ is the $4\times4$ Hadamard matrix.

   For general parameters $M_{i}(i=8,...,15)$, it is not obvious how to remove $\varphi_{j}$ from (\ref{4}) by maximization. Based on the symmetric consideration, we assume
   \begin{eqnarray}
    M_{9}=M_{10}=M_{11}=M_{12}=M_{13}=M_{14}.\label{5y}
   \end{eqnarray}
   Although the assumption may limit the entanglement detecting power of the optimal EW derived from operator $\hat{M}$. It greatly simplifies the analysis. The assumption (\ref{5y}) on the parameters $M_{i}$ simplifies $g_{1}(\varphi_{+},\varphi_{-})$ to
\begin{equation}\label{6}\nonumber
g_{1}=\sqrt{(A_{1}c_{+}+A_{2}c_{-})^2+(A_{5}s_{+})^2}+|A_{2}c_{+}+A_{4}c_{-}|.
   \end{equation}
It is not difficult to show that the second derivative of $g_{1}$ with respect to $c_{-}$ is always non negative. Hence we have
\begin{eqnarray}\label{7}\nonumber
g_{2}(\varphi_{+})=\max_{c_{-}}g_{1}(\varphi_{+},\varphi_{-})=\max\{g_{1}|_{c_{-}=+1},g_{1}|_{c_{-}=-1}\}\\
=\sqrt{(A_{1}c_{+}+ A_{2})^2+(A_{5}s_{+})^2}+|A_{2}c_{+}+A_{4}|.
\end{eqnarray}
In the last equality, we have merged the $c_{-}=\pm1$ cases into the sign of $c_{+}$. Thus $\tilde{g}=\max_{\varphi_{+}}g_{2}(\varphi_{+})$. The maximization over $\varphi_{+}$ can be carried out and we at last have
\begin{equation}\label{8}
  \tilde{g}=\left\{\begin{array}{l}
                   \text{sign}(M_{9})\frac{9M_{9}^2-M_{8}M_{15}}{6M_{9}-M_{8}-M_{15}}, \text{\quad  if\quad} (\frac{M_{8}}{M_{9}},\frac{M_{15}}{M_{9}})\in \Delta; \\
                   \max\{\frac{1}{2}(|M_{8}+M_{9}|+|M_{8}-M_{9}|), \frac{1}{2}(|M_{15}+M_{9}|\\
                    +|M_{15}-M_{9}|)\}, \text{\quad otherwise}.
                 \end{array}
                 \right.
\end{equation}
Where $\Delta$ is a region shown in Fig.1.

Denote $f_{1}(\boldsymbol{\theta})=\max_{\boldsymbol{\varphi}}f(\boldsymbol{\theta},\boldsymbol{\varphi})$. by optimizing on $\theta_{4}$ and denoting $f_{2}(\theta_{1},\theta_{2},\theta_{3})=\max_{\theta_{4}}f_{1}(\boldsymbol{\theta})$, we then have
\begin{eqnarray}\label{9}\nonumber
  f_{2}(\theta_{1},\theta_{2},\theta_{3})=M_{6}z_{1}z_{2}+M_{5}z_{1}z_{3}+M_{3}z_{2}z_{3}+\\
\sqrt{[M_{4}z_{1}+M_{2}z_{2}+M_{1}z_{3}+M_{7}z_{1}z_{2}z_{3}]^2+(\tilde{g}t_{1}t_{2}t_{3})^2}.
\end{eqnarray}

The maximal mean of the operator $\hat{M}$ over all product states is denoted as $\Lambda=\max_{|\psi\rangle}\langle\psi|\hat{M}|\psi\rangle=\max_{\boldsymbol{\theta},\boldsymbol{\varphi}}f(\boldsymbol{\theta},\boldsymbol{\varphi})=\max_{\theta_{1},\theta_{2},\theta_{3}}f_{2}(\theta_{1},\theta_{2},\theta_{3})$.
We have the following lemma:
\begin{lemma}
In parameter space $(M_{1},M_{2},M_{3},M_{4},M_{5}$ ,$M_{6}, M_{7})/\widetilde{g}$, there are  points $P_{i}(i=1,...,8)$ with coordinates be $(j_{1},j_{2},-j_{1}j_{2},j_{3},-j_{1}j_{3},-j_{2}j_{3},j_{1}j_{2}j_{3})$ subjected to $|j_{1}|=|j_{2}|=|j_{3}|=1$. Then inside the polyhedron $(P_{1},P_{2},P_{3},P_{4},P_{5},P_{6},P_{7},P_{8})$ we have
\begin{equation}\label{9a}
  \Lambda=\widetilde{g}.
\end{equation}
\end{lemma}
Proof: For each point $P_{i}$, a direct calculation shows that $\Lambda(P_{i})=\widetilde{g}$. Notice that $f_{1}$ is linear with respect to the parameters $M_{i},(i=1,...,7)$. So inside the polyhedron $(P_{1},P_{2},P_{3},P_{4},P_{5},P_{6},P_{7},P_{8})$, we have $f_{1}(M_{1},M_{2},M_{3},M_{4},M_{5},M_{6},M_{7})=\sum_{i=1}^8p_{i}f_{1}(P_{i})$ with some probability distribution $\{p_{i}\}$. Then we have $\Lambda(M_{1},M_{2},M_{3},M_{4},M_{5},M_{6},M_{7})\leq \sum_{i=1}^8p_{i}\Lambda(P_{i})=\widetilde{g}$ due to the fact that each $\Lambda(P_{i})$ may be achieved at its own special $\theta_{j}$ variables. On the other hand, we get the lower bound of  $\Lambda(M_{1},M_{2},M_{3},M_{4},M_{5},M_{6},M_{7})$ by noticing that $f_{2}(\theta_{1}=\pm\frac{\pi}{2},\theta_{2}=\pm\frac{\pi}{2},\theta_{3}=\pm\frac{\pi}{2})=\widetilde{g}$, so that $\Lambda(M_{1},M_{2},M_{3},M_{4},M_{5},M_{6},M_{7})\geq \widetilde{g}$. Thus (\ref{9a}) follows.

In the outside of the polyhedron, we can have $\Lambda>\widetilde{g}$.

\section{Matched entanglement witness and separable criterion}

   A four qubit GHZ diagonal state can be written as
   \begin{eqnarray}\label{11}\nonumber
 \rho&=&\frac{1}{16}(IIII+R_{1}IIZZ+R_{2}IZIZ+R_{3}IZZI\\
&&+R_{4}ZIIZ+R_{5}ZIZI+R_{6}ZZII+R_{7}ZZZZ\nonumber\\
&&+R_{8}XXXX+R_{9}XXYY+R_{10}XYXY\nonumber\\
&&+R_{11}XYYX+R_{12}YXXY+R_{13}YXYX\nonumber \\
&&+R_{14}YYXX+R_{15}YYYY).
\end{eqnarray}
   Then $Tr\rho\hat{M}=\sum_{i=1}^{15}M_{i}R_{i}$. Let
   \begin{equation}\label{12}
     \mathcal{L}=\frac{\Lambda}{\sum_{i=1}^{15}M_{i}R_{i}}.
   \end{equation}
   With the convention of $\sum_{i=1}^{15}M_{i}R_{i}>0$, we say the entanglement of $\rho$ is detected if $\mathcal{L}<1$. For all possible optimal EWs,
   we want to find an EW with the smallest $\mathcal{L}$. We will call it matched EW with respect to the given state $\rho$. Hence the problem is to minimize $\mathcal{L}$ with respect to $\hat{M}$.
   \begin{equation}\label{13}
     \mathcal{L}_{\min}=\min_{\hat{M}}\mathcal{L}.
   \end{equation}

\subsection{Matching anti-diagonal elements of states}

   In order to minimize $\mathcal{L}$ with respect to $M_{i}$ (i=1,...,15), we first consider
     $\tilde{R}=\max\sum_{i=8,...,15}M_{i}R_{i}/\tilde{g}$. Here the anti-diagonal part of state $\rho$ in (\ref{11}) is described by $R_{i}$ (i=8,...,15). With the assumption (\ref{5y}) and notations $x=\frac{M_{8}}{M_{9}},y=\frac{M_{15}}{M_{9}}$, we have $\tilde{R}=\max_{x,y}\frac{1}{\mathcal{L}_{g}},$ where

     \begin{equation}\label{14}
          \mathcal{L}_{g}=\frac{\tilde{g}}{M_{9}(xR_{8}+yR_{15}+R'_{9})}.
     \end{equation}
   With $R'_{9}=\sum_{i=9}^{14}R_{i}$. Suppose $R'_{9}>0$ without loss of generality, we may choose $M_{9}>0$ to match with the sign of $R'_{9}$. If $R_{8}<0$, we consider the region in the left side of Fig.1 specified  by its boundary lines $A'A''$ , $D'D''$ and curve $A'D'$. In this region, we have $\tilde{g}=|M_{8}|=-x|M_{9}|$, then
   \begin{equation}\label{15}
          \mathcal{L}^{-1}_{g}=-R_{8}+\frac{y}{|x|}R_{15}+\frac{1}{|x|}R'_{9}.
     \end{equation}
   For a given $y$, the maximum of $\mathcal{L}^{-1}_{g}$ achieves at the region boundary lines $A'A''$ , $D'D''$ and curve $A'D'$. $\mathcal{L}^{-1}_{g}$ monotonically increases with the decrease of $|x|$ since $R'_{9}+yR_{15}>0$ (we may choose the sign of $y$ to be the sign of $R_{15}$. We omit the case $R'_{9}+yR_{15}<0$ which gives rise to a local maximum $-R_{8}$ for $\mathcal{L}^{-1}_{g}$ when $x\rightarrow -\infty$). Along with the lines $A'A''$ and $D'D''$, $\mathcal{L}^{-1}_{g}$ monotonically increases with the decrease of $|x|$. So, we only need to consider the maximization of $\mathcal{L}^{-1}_{g}$ on curve $A'D'$. Similar analyses can be applied to the other regions in Fig.1 except region $\Delta$. Thus the region for maximization of $\mathcal{L}^{-1}_{g}$ can be reduced to $\Delta$ including its boundary. In region $\Delta$, we have
   \begin{equation}\label{16}
     \mathcal{L}^{-1}_{g}=\frac{(6-x-y)(xR_{8}+yR_{15}+R'_{9})}{9-xy}.
   \end{equation}
   We consider the change of $\mathcal{L}^{-1}_{g}$ on the straight line connecting point $C'$ and some point in curve $A'D'$ or curve $A'B'$. The equation for such a straight line is $y=kx+3(1-k)$. On the line we have $\frac{(6-x-y)}{9-xy}=\frac{1+k}{3+kx}$, thus
   \begin{equation}\label{17}
     \mathcal{L}^{-1}_{g}=(1+k)(\frac{R_{8}}{k}+R_{15})+\frac{3(1+k)}{3+kx}(\frac{R'_{9}}{3}-\frac{R_{8}}{k}-kR_{15}).
   \end{equation}
   Notice that $k\geq0$ and in the domain $\Delta$ we can verify that $3+kx>0$. Hence $\mathcal{L}^{-1}_{g}$ maximizes at point $C'$ when
   \begin{equation}\label{18}
    \frac{R_{8}}{k}+kR_{15}>\frac{R'_{9}}{3}.
   \end{equation}
   and maximizes at curve $B'A'D'$ otherwise. A tighter alternative of (\ref{18}) is
   \begin{equation}\label{19}
   \frac{R_{8}R_{15}}{R'^2_{9}}>\frac{1}{36}.
   \end{equation}

   We further consider the maximization of $\mathcal{L}^{-1}_{g}$ on curve $A'D'$. We have the equation of curve $A'D'$ to be $\frac{(6-x-y)}{9-xy}=-\frac{1}{x}$ from (\ref{8}), hence
   \begin{equation}\label{20}
     \mathcal{L}^{-1}_{g}=-R_{8}+\frac{1}{2}R_{15}-(3R_{15}+R'_{9})\frac{1}{x}-\frac{9R_{15}}{2}\frac{1}{x^2}.
   \end{equation}
   So $\mathcal{L}^{-1}_{g}$ achieves its maximum in curve $A'D'$ with $x=-\frac{9R_{15}}{3R_{15}+R'_{9}}$ when
   \begin{equation}\label{21}
     \frac{R_{15}}{R'_{9}}>\frac{1}{6}.
   \end{equation}
   It achieves its maximum at point $A'$ otherwise. Similarly, $\mathcal{L}^{-1}_{g}$ achieves its maximum in curve $A'B'$ with $y=-\frac{9R_{8}}{3R_{8}+R'_{9}}$ when
   \begin{equation}\label{22}
     \frac{R_{8}}{R'_{9}}>\frac{1}{6}.
   \end{equation}
   It achieves its maximum at point $A'$ otherwise. The explicit formula for $\tilde{R}$ (regardless of the sign of $R'_{9}$) is
\begin{equation}\label{23}
  \tilde{R}=\left\{\begin{array}{l}
  |R'_{9}-R_{8}-R_{15}|, \text{  if  } \frac{R_{8}}{R'_{9}}\leq\frac{1}{6},\frac{R_{15}}{R'_{9}}\leq\frac{1}{6};\\
  |\frac{1}{3}R'_{9}+R_{8}+R_{15}|,  \text{  if  } \frac{R_{8}R_{15}}{R'^2_{9}}\geq\frac{1}{36};\frac{R_{8}}{R'_{9}}>0;\frac{R_{15}}{R'_{9}}>0 \\
  |R_{15}-R_{8}+\frac{1}{3}R'_{9}+\frac{R'^2_{9}}{18R_{15}}|,\\
  \text{\quad \quad \quad \quad \quad \quad if  } \frac{R_{15}}{R'_{9}}>\frac{1}{6}  , \frac{R_{8}R_{15}}{R'^2_{9}}<\frac{1}{36};\\
  |R_{8}-R_{15}+\frac{1}{3}R'_{9}+\frac{R'^2_{9}}{18R_{8}}|,\\
  \text{\quad \quad \quad \quad \quad \quad if  } \frac{R_{8}}{R'_{9}}>\frac{1}{6} , \frac{R_{8}R_{15}}{R'^2_{9}}<\frac{1}{36}.
  \end{array}
  \right.
\end{equation}

\subsection{Matching diagonal elements of state}

The quantity $\mathcal{L}_{\min}$ defined in (\ref{13}) is
\begin{equation}\label{24}
  \mathcal{L}_{\min}=\min_{M_{m},m=1,...,7}\frac{\Lambda}{\sum_{m=1}^{7}M_{m}R_{m}+\widetilde{g}\widetilde{R}}\\
\end{equation}

The parameter space $(M_{1},M_{2},M_{3},M_{4},M_{5},M_{6},M_{7})/\widetilde{g}$ is divided into two parts, one is the outside of polyhedron $(P_{1},P_{2},P_{3},P_{4},P_{5},P_{6},P_{7},P_{8})$, the other is the inside of the polyhedron (including the boundary). Suppose the point $(N_{1},N_{2},N_{3},N_{4},N_{5},N_{6},N_{7})/\widetilde{g}$ be on the boundary of the polyhedron, then
\begin{equation}\label{25}
\mathcal{L}_{boundary}=[(\sum_{m=1}^{7}N_{m}R_{m})/\widetilde{g}+\widetilde{R}]^{-1}.
\end{equation}

Notice that the origin is inside the polyhedron since it is the geometrical center of the points $P_{1},...,P_{8}$. Let $\delta>0$, then $(M_{1},M_{2},M_{3},M_{4},M_{5},M_{6},M_{7})\widetilde{g}^{-1}=(1+\delta)(N_{1},N_{2},N_{3},N_{4},N_{5},N_{6},N_{7})\widetilde{g}^{-1} $ be a point outside the polyhedron in the parameter space, we have $\Lambda=(1+\delta)\widetilde{g}$, $\mathcal{L}=[(\sum_{m=1}^{7}N_{m}R_{m})/\widetilde{g}+\widetilde{R}/(1+\delta)]^{-1}>\mathcal{L}_{boundary}$. Hence the minimal $\mathcal{L}$ can only be achieved inside the polyhedron.

Let $(M_{1},M_{2},M_{3},M_{4},M_{5},M_{6},M_{7})\widetilde{g}^{-1}=\delta(N_{1},N_{2},N_{3}$, $N_{4},N_{5},N_{6},N_{7})\widetilde{g}^{-1} (1\geq\delta\geq0)$ be a point inside the polyhedron in the parameter space, we have $\Lambda=\widetilde{g}$, thus $\mathcal{L}=[\delta(\sum_{m=1}^{7}N_{m}R_{m})/\widetilde{g}+\widetilde{R}]^{-1}>\mathcal{L}_{boundary}$ as far as $\sum_{m=1}^{7}N_{m}R_{m}$ is positive. If $\sum_{m=1}^{7}N_{m}R_{m}$ is negative, we have $\mathcal{L}=\widetilde{R}^{-1}=\mathcal{L}_{origin}$. Thus the minimal $\mathcal{L}$ is achieved either on the boundary or on the original point of the parameter space. Notice that $\mathcal{L}_{boundary}^{-1}$ is linear with respect to the parameter $(N_{1},N_{2},N_{3},N_{4},N_{5},N_{6},N_{7})\widetilde{g}^{-1}$, hence the maximal of $\mathcal{L}_{boundary}^{-1}$ is achieved on one of the extremal points $P_{i},i=1,...,8$. The original point is excluded since a point with non-negative $\sum_{m=1}^{7}N_{m}R_{m}$ always exists on the boundary. In fact, the values of $\sum_{m=1}^{7}N_{m}R_{m}$ for points $P_{i}, (i=1,...,8)$ are $1-8(\rho_{i,i}+\rho_{17-i,17-i}) (i=1,...,8)$, respectively. Then we choose the maximum of $1-8(\rho_{i,i}+\rho_{17-i,17-i})$ to minimize $\mathcal{L}_{boundary}$. Thus
\begin{equation}\label{26}
  \mathcal{L}_{\min}=[1-8\min_{i}(\rho_{i,i}+\rho_{17-i,17-i})+\widetilde{R}]^{-1}.
\end{equation}

\subsection{Separable criteria}
The separable criterion is $\mathcal{L}_{\min}\geq 1 $. Namely,
\begin{equation}\label{27}
  \min_{i}(\rho_{i,i}+\rho_{17-i,17-i})\geq \frac{1}{8}\tilde{R},
\end{equation}
The separable criteria have an operational meaning of comparing the anti-diagonal part of the density matrix (represented by $\tilde{R}$) and the diagonal part of the density matrix (represented the minimal $\rho_{i,i}+\rho_{17-i,17-i}$). A state is necessarily separable when its diagonal part is larger than its anti-diagonal part and it is entangled otherwise.

The simplest separable criterion comes from the first line of (\ref{23}) and  (\ref{27}). It is
\begin{equation}\label{28}
 \text{Criterion I}: |\rho_{1,16}|\leq\frac{1}{2}\min_{i}(\rho_{i,i}+\rho_{17-i,17-i}),
\end{equation}
Criterion I is just the criterion for the generalized Werner states. In fact it is just the positive partial transpose fully separable criterion for four qubit GHZ diagonal states.
 We may substitute the second lines of (\ref{23}) and into (\ref{27}) to obtain the following criterion.
\begin{equation}\label{29}
\text{Criterion II}:\frac{1}{3}|\rho_{4,13}+\rho_{6,11}+\rho_{7,10}|\leq \Omega,
\end{equation}
Where
\begin{equation}\nonumber
\Omega=\frac{1}{2}\min_{i}(\rho_{i,i}+\rho_{17-i,17-i})\\.
\end{equation}

The other criteria are
\begin{equation}\label{30}
\text{Criterion III:   }|R_{15}-R_{8}+\frac{1}{3}R'_{9}+\frac{R'^2_{9}}{18R_{15}}|\leq 16\Omega,
\end{equation}
 and
\begin{equation}\label{31}
\text{Criterion IV:   }|R_{8}-R_{15}+\frac{1}{3}R'_{9}+\frac{R'^2_{9}}{18R_{8}}|\leq 16\Omega.
\end{equation}
We have classified the separable criteria into four types for the full separability of four qubit GHZ diagonal states. We will show that two of them are necessary and sufficient for the full separability of four qubit (highly) symmetric GHZ diagonal states.

\section{Application to highly symmetric GHZ diagonal states}

A four qubit highly symmetric GHZ diagonal state $\rho$ is the mixture of GHZ basis states with the probabilities:\{$ p_{i},i=1,...,16$\} and $p_{2}=p_{3}=...=p_{8}$; $p_{9}=p_{10}=...=p_{15}$.
The state is symmetric under interchange of any pair of qubits.
 Hence we get four positive parameters $p_{1},p_{2},p_{15},p_{16}$ with normalization
\begin{equation}\label{32}
  p_{1}+p_{16}+7(p_{2}+p_{15})=1.
\end{equation}
The nonzero entries of $\rho$ are $\rho_{1,1}=\rho_{16,16}=\frac{1}{2}(p_{1}+p_{16})$;
$\rho_{1,16}=\rho_{16,1}=\frac{1}{2}(p_{1}-p_{16})$; $\rho_{2,2}=\rho_{3,3}=...=\rho_{15,15}=\frac{1}{2}(p_{2}+p_{15})$;
$\rho_{2,15}=\rho_{3,14}=...=\rho_{15,2}=\frac{1}{2}(p_{2}-p_{15})$.

We have numerically calculated the boundaries of separable state sets for above states
in the cases of $p_{16}=0$, and $p_{16}=0.3$. We choose $p_{2}$ and $p_{15}$ as free parameters
and $p_{1}$ is determined by the normalization (\ref{32}). The boundaries are shown in Fig.2.
The numeric calculation has rounds of three steps: (i) to choose $M_{i}$ randomly (ii) to calculate $\Lambda$ (iii)to record the minimal $\mathcal{L}$.

For the convenience of later uses, we list the relevant $R_{j}$ below
\begin{eqnarray}\nonumber
  R_{7} &=&R_{1}=1-8(p_{2}+p_{15}), \\
  R_{8} &=& 1-2p_{16}-14p_{15}, \nonumber\\
  R_{15}&=&-R_{9}=1-2p_{16}-8p_{2}-6p_{15}, \nonumber
\end{eqnarray}
 $R'_{9}=6R_{9}$, $R'_{1}=6R_{1}$. We alternatively use parameters $u,v,\alpha$  for convenience, with $u=1-2p_{16}$,  $(p_{15},p_{2})=(\frac{vu}{\alpha},\frac{(1-v)u}{\alpha}), v\in[0,1]$. Hence
    \begin{eqnarray}
    R_{8}=u(1-\frac{14v}{\alpha}), \label{32a}\\
    R_{15}=u(1-\frac{8-2v}{\alpha}).\label{32b}
    \end{eqnarray}

\subsection{Necessary criteria}

    For our four qubit highly symmetric GHZ diagonal state $\rho$, we have
\begin{equation}\nonumber
      \Omega=\min(\rho_{1,1},\rho_{2,2}).
\end{equation}
Criterion I turns out to be
    \begin{equation}\label{33}
      |p_{1}-p_{16}|\leq (p_{2}+p_{15})\\
    \end{equation}
    when $p_{1}+p_{16}\geq(p_{2}+p_{15})$.
    When $p_{1}>p_{16}$, the criterion gives the upper bound of $\alpha$ corresponding to straight line boundaries $GH$, $PQ$ and $KL$ in Fig.2, with $\alpha=8$. We have denoted the intersections of the criteria by points $G,H...$ in the Fig.2. When $p_{1}<p_{16}$, the criterion gives the lower bound of $\alpha$ corresponding to straight line boundary $MN$ in Fig.2, with $\alpha=6$.

    Criterion II now is
    \begin{equation}\label{34}
      |p_{2}-p_{15}|\leq p_{1}+p_{16}.
    \end{equation}
    It is
    \begin{eqnarray}
      \alpha\geq u(8-2v) \text{\quad for } v\leq\frac{1}{2} \label{35}\\
      \alpha\geq u(6+2v) \text{\quad for } v\geq\frac{1}{2} \label{36}.
    \end{eqnarray}
    Condition (\ref{35}) accounts for the line boundary $GJ$ in Fig.2.

    Criterion III turns out to be $|p_{2}-p_{15}|\leq p_{1}+p_{16}$ too.

     When $p_{1}+p_{16}\geq p_{2}+p_{15}$, criterion IV leads to
    \begin{eqnarray}
      v \geq\frac{1}{36}[18-\alpha-\sqrt{(\alpha+42)(\alpha-6)}], \label{37} \\
      v \leq\frac{1}{36}[32-\alpha+\sqrt{(56-\alpha)(8-\alpha)}], \label{38}
    \end{eqnarray}
with $\alpha\in[6,8]$. Inequality (\ref{37}) gives rise to boundary curve $KN$ in Fig.2, with
\begin{equation}\label{39}
      v_{K}=0;  \text{\quad for \quad} \alpha=8; \text{\quad}   v_{N}=\frac{1}{3} \text{\quad for\quad} \alpha=6.
\end{equation}
Inequality (\ref{38}) gives rise to boundary curve $LM$ in Fig.2. with
\begin{equation}\label{40}
      v_{M}=1;  \text{\quad for \quad} \alpha=6; \text{\quad}   v_{L}=\frac{2}{3} \text{\quad for\quad} \alpha=8.
\end{equation}
The straight line $KL$ then is limited to $v\in [v_{K},v_{L}]$ and the straight line $MN$ is limited to $v\in [v_{N},v_{M}]$.

When $p_{1}+p_{16}\leq p_{2}+p_{15}$, criterion IV leads to
\begin{eqnarray}
  \alpha &\geq& a_{1}+\sqrt{a_{1}^2+b_{1}}   \text{ for } v\in [v_{1},v_{2}],   \label{41}\\
  \alpha &\geq& a_{2}-\sqrt{a_{2}^2-b_{2}}   \text{ for } v\in [v_{3},v_{4}],   \label{42}
\end{eqnarray}
where $a_{1}=3u+(7+u)v, b_{1}=4u(4-37v+9v^2)$, $a_{2}=4u+(7-u)v, b_{2}=4u(3v+2)^2$. Here $v_{1}=\frac{1}{18}(9-4u-\sqrt{(4u+21)(4u-3)})$ and $v_{4}=\frac{2}{9}(4-u-\sqrt{(7-u)(1-u)})$ are determined by condition $p_{1}+p_{16}=p_{2}+p_{15}$ (namely $\alpha=8u$) and the equality in (\ref{37}) and (\ref{38}), respectively.  $v_{2}=\frac{5u-2}{3(1+u)}$ and $v_{3}=\frac{4-3u}{1+u}$ come from physical boundary condition $p_{16}+p_{2}+p_{15}=1$ (namely $\alpha=\frac{14u}{1+u}$) and the equality in (\ref{41}) and (\ref{42})), respectively.

It happens that the straight line $\alpha=u(6+2v)$ and the curve $\alpha= a_{2}-\sqrt{a_{2}^2-b_{2}}$ intersect at $(v=v_{3}, \alpha=\frac{14u}{1+u})$. We can verified that $a_{2}-\sqrt{a_{2}^2-b_{2}}>u(6+2v)$ when $v>v_{3}$. Thus the inequality (\ref{42}) is better than inequality (\ref{36}) as a separable criterion. We can also show that $a_{1}+\sqrt{a_{1}^2+b_{1}}>u(8-2v)$, thus inequality (\ref{41}) is better than inequality (\ref{35}) as a separable criterion. We conclude that criterion II and criterion III are not useful in determining the separable boundary of four qubit highly symmetric GHZ diagonal state set. Criterion I and criterion IV suffice as necessary criterion set for full separability.

 When $p_{16}=0$. Inequality (\ref{41}) turns out to be (\ref{35}) and inequality (\ref{42}) leads to
 \begin{equation}
      \alpha = 4+6v, \text{\quad for  } v\in [\frac{1}{2},\frac{2}{3}] , \label{43}
 \end{equation}
Equation (\ref{43}) is just the straight line boundary $HJ$ in Fig.2. As a criterion, it is better than inequality (\ref{36}). The states inside the triangle $GHJ$ are separable due to the convexity of separable state set.

We have shown the boundaries of fully separable states for $p_{16}=0$ and $p_{16}=0.3$ in Fig.2, respectively. The figure for the case of $p_{16}\in (0,\frac{1}{8})$ is quite different. As shown in Fig.2 for $p_{16}=0.0625$, the fully separable state set has boundary $PQSTUVP$. Straight line $PQ$ is specified by $\alpha=8, v\in [0,\frac{2}{3}]$. Curve $PV$ ($QS$) is determined by the equality in (\ref{37})((\ref{38})) until it reaches $\alpha=8u$ with $v=v_{1}(v_{4})$. Curve $VU$ ($ST$) is determined by the equality in (\ref{41})((\ref{42})) until it reaches $\alpha=\frac{14u}{1+u}$ with $v=v_{2}(v_{3})$. Finally the straight line section $UT$ represents the physical boundary $p_{1}=0$.
\begin{figure}[tpb]
\includegraphics[ trim=0.000000in 0.000000in -0.138042in 0.000000in,
height=2.5in, width=3.5in]{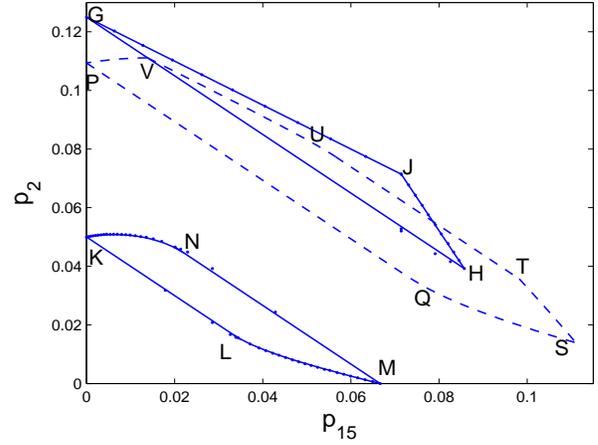}
\caption{The numeric calculated boundaries of fully separable state set of four qubit highly symmetric GHZ diagonal states with $p_{16}=0$(GJHG) and $p_{16}=0.3$(KLMNK) specified by dots. Theoretical results for the boundaries are displayed with solid lines. The dashed lines are for the theoretical results of $p_{16}=0.0625$. Criterion I accounts for straight lines $GH$, $KL$, $MN$ and $PQ$. Criterion IV accounts for straight lines $GJ$, $JH$ and curves $KN,LM$, $QS$, $ST$,$PV$ and $UV$. The straight line $UT$ is due to physical condition $p_{1}=0$.
}
\end{figure}
    \subsection{Sufficient criteria}

    The sufficient condition of separability relies on the ability of decomposing the state into probability
    mixture of product states. Usually, it is rather technical to write down the decomposition. For the known operator $\hat{M}$, we will find the product state corresponding to its largest mean $\Lambda$ and use it to construct the explicit decomposition of a state $\rho$ at the boundary of fully separable state set.
    \subsubsection{Sufficiency of criterion I}
    In the case of Criterion I, we start by setting $M_{1}=M_{2}=...=M_{6}=0,M_{7}=1, M_{8}=M_{15}=-M_{9}=\pm1$. We thus have the operator $\hat{M}=ZZZZ\pm(XXXX-XXYY-XYXY-XYYX-YXXY-YXYX-YYXX+YYYY)$. The EW is $\mathbb{I}-\hat{M}$. The maximal mean of $\hat{M}$ over product state $|\psi\rangle=\bigotimes_{i=1}^{4}|\psi_{i}\rangle$ is
    \begin{equation}\label{44}
      \Lambda=\max_{\boldsymbol{\theta},\boldsymbol{\varphi}}[\prod_{i=1}^{4}z_{i}\pm\prod_{i=1}^{4}t_{i}\cos(\sum_{j=1}^{4}\varphi_{j})]
    \end{equation}
    The maximum is known to be 1 which is achieved when (i) $\prod_{i=1}^{4}z_{i}=1$ or (ii) $\prod_{i=1}^{4}t_{i}=\pm1$ and $\sum_{j=1}^{4}\varphi_{j}=0$. Case (i) corresponds to separable states $|0000\rangle, |0011\rangle,|0101\rangle, |1001\rangle, |0110\rangle,|1010\rangle,|1100\rangle, |1111\rangle $. Case (ii) is realized by separable state
    \begin{equation}\nonumber
      |\psi(\varphi_{1},\varphi_{2},\varphi_{3})\rangle=\frac{1}{4}[\bigotimes_{j=1}^{3}(|0\rangle+e^{i\varphi_{j}}|1\rangle)](|0\rangle\pm e^{-i\Sigma_{j=1}^{3}\varphi_{j}}|1\rangle).
    \end{equation}
    In order to construct the GHZ diagonal states, we may expand the product state $|\psi(\varphi_{1},\varphi_{2},\varphi_{3})\rangle$
    with products of Pauli matrices. Then we will eliminate the unnecessary terms by the following procedure.
    Denote $\varrho_{1}(\varphi_{1},\varphi_{2},\varphi_{3})=|\psi(\varphi_{1},\varphi_{2},\varphi_{3})\rangle\langle\psi(\varphi_{1},\varphi_{2},\varphi_{3})|$. Let $\boldsymbol{k}=(k_{1},k_{2},k_{3})\in\{0,1\}^{\otimes3}$, denote
    \begin{eqnarray}\nonumber
    \varrho_2(\varphi_{1},\varphi_{2},\varphi_{3})=\frac{1}{8}\Sigma_{\boldsymbol{k}}\varrho_{1}(\varphi_{1}+k_{1}\pi,\varphi_{2}+k_{2}\pi,\varphi_{3}+k_{3}\pi) \\
    \varrho_3(\varphi)=\frac{1}{2}[\varrho_2(\varphi,\varphi,\varphi)+\varrho_{2}^{*}(\varphi,\varphi,\varphi)]\nonumber
    \end{eqnarray}
    The density matrix $\varrho_2(\varphi_{1},\varphi_{2},\varphi_{3})$ is already in 'X' shape, all its entries are nullified except diagonal and anti-diagonal entries. The real density matrix $\varrho_3(\varphi)$ is
    \begin{eqnarray} \nonumber
    \frac{1}{16}IIII\pm\frac{1}{16}[\cos3\varphi\cos^3\varphi XXXX-\sin3\varphi\sin^3\varphi YYYY\nonumber\\
    +\cos3\varphi\cos\varphi\sin^2\varphi(XYYX+YXYX+YYXX)\nonumber\\
    -\sin3\varphi\sin\varphi\cos^2\varphi(XXYY+XYXY+YXXY)]\nonumber
    \end{eqnarray}
    The permutational symmetry of the state requires $\cos3\varphi\cos\varphi\sin^2\varphi=-\sin3\varphi\sin\varphi\cos^2\varphi$. It leads to $\varphi=\frac{i\pi}{4}$, with $i=0,...,7$. We obtain independent separable states $\varrho_3(0)$, $\varrho_3(\frac{\pi}{4})$, $\varrho_3(\frac{\pi}{2})$. A mixture of these states will give rise to
    \begin{equation}\nonumber
    \varrho_4=(1-q_{1}-q_{2})\varrho_3(0)+q_{1}\varrho_3(\frac{\pi}{4})+q_{2}\varrho_3(\frac{\pi}{2}).
    \end{equation}
    Where $0\leq q_{1},q_{2}\leq 1; q_{1}+q_{2}\leq 1$. Then a state on line sections $GH$,$KL$  $PQ$  in Fig.2 can be written as
    \begin{equation}\label{45}
      \rho=u\varrho_{4}+p_{16}(|0000\rangle\langle0000|+|1111\rangle\langle1111|),
    \end{equation}
    where $`+$' is chosen from $`\pm$' for state $\varrho_{4}$. We then have
    \begin{equation}\label{45a}
    R_{8}=u(1-\frac{5}{4}q_{1}-q_{2}), R_{9}=u(-\frac{1}{4}q_{1}),R_{15}=u(q_{2}-\frac{1}{4}q_{1}).
    \end{equation}
Noti ce that $R_{15}=-R_{9}$, so $q_{2}=\frac{1}{2}q_{1}$. Hence $q_{1}\leq \frac{2}{3}$. Comparing $R_{i}$ with their expressions of parameters ($\alpha=8, v$). We have $v=q_{1}$,
    \begin{equation}\label{46}
      0\leq v\leq \frac{2}{3},
    \end{equation}
    for the states on the line sections $GH$ and $KL$.

      A state on line section $MN$  in Fig.2 can be written as
    \begin{equation}\label{47}
      \rho=(1-2p_{1})\varrho_{4}+p_{1}(|0000\rangle\langle0000|+|1111\rangle\langle1111|),
    \end{equation}
    where $`-$' is chosen from $`\pm$' for state $\varrho_{4}$. We compare $R_{i}$ obtained from (\ref{47}) with their expressions of parameters ($\alpha=6, v$). We have $v=1-q_{1}$,
   \begin{equation}\label{48}
      \frac{1}{3}\leq v\leq 1,
    \end{equation}
    for the states on the line section $MN$.

 The sufficient conditions for the full separability of the states on straight line sections $KL,MN,GH$ are proven. The sufficient conditions of criterion I coincide with the necessary conditions. So criterion I is necessary and sufficient for the full separability of states on straight line sections $KL,MN,GH$.

 The general Werner state is a special highly symmetric GHZ diagonal state, with $p_{2}=...=p_{15}=p_{16}$ and  $p=p_{1}-p_{2}>0$. Criterion I reads $p\leq \frac{1}{9}$. It is the necessary and sufficient criterion of full separability for the general Werner states\cite{Chen2015}\cite{GuhneSeevinck}. Thus the criterion set $\mathcal{C}_{1}$=\{Criterion I\} is necessary and sufficient for the full separability of the states in the general Werner state set $\mathcal{S}_{1}$.

\subsubsection{Sufficiency of criterion IV}
   The solutions to the maximization of (\ref{7}) are $s_{+}=0$ and $c_{+}=\frac{A_{2}}{1-A_{1}}=\frac{M_{8}-M_{15}}{6M_{9}-M_{8}-M_{15}}$. The later leads to the first line of (\ref{8}) for $\tilde{g}$. The corresponding separable state can be written as
   \begin{equation}\nonumber
      |\psi(\varphi,\boldsymbol{m})\rangle=\frac{1}{4}\bigotimes_{j=1}^{4}(|0\rangle+e^{i\varphi_{j}}|1\rangle).
    \end{equation}
 where $\boldsymbol{m}=(m_{1},m_{2},m_{3},m_{4})\in\{0,1\}^{\otimes4}$, $\varphi_{j}=\varphi+m_{j}\pi$. Let $l=\mod(\sum_{j}m_{j},2)$ be the parity of $\boldsymbol{m}$. Define $\varrho_{+(-)}(\varphi)=\sum_{\boldsymbol{m}|l=0(1)}|\psi(\varphi,\boldsymbol{m})\rangle \langle\psi(\varphi,\boldsymbol{m})|$. Then we have
 \begin{eqnarray} \label{48a}
   \varrho_{\pm}(\varphi)=\frac{1}{16}[IIII\pm(\cos^4\varphi XXXX+\sin^4\varphi YYYY\nonumber\\
   +\cos^2\varphi\sin^2\varphi(XXYY+XYXY+XYYX\nonumber\\
   +YXXY+YXYX+YYXX))].
 \end{eqnarray}
 Denote
 \begin{equation}\nonumber
   \varrho_{5\pm}=\frac{1}{1+\sin^2\varphi}(\varrho_{\pm}(\varphi)+\sin^2\varphi\varrho_{\mp}(\frac{\pi}{2})).
 \end{equation}
 The separable states in the curves $LM$ and $KN$ in Fig.2 can be expressed as
 \begin{equation}\label{49}
   \rho=(1-R_{7})\varrho_{5\mp}+\frac{R_{7}}{2}(|0000\rangle\langle0000|+|1111\rangle\langle1111|),
 \end{equation}
respectively. Hence $R_{8}=\mp K\cos^{4}\varphi, R_{15}=\pm K\cos^{2}\varphi\sin^{2}\varphi$ with $K=\frac{1-|R_{7}|}{1+\sin^{2}\varphi}=\frac{8u}{\alpha(1+\sin^{2}\varphi)}$. Compare them with equations (\ref{32a}) (\ref{32b}), we have
\begin{equation}\label{50}
  v=\frac{1}{1+\sin^{2}\varphi}, \text{  }
  \alpha=\frac{14-8\cos^{4}\varphi}{1+\sin^{2}\varphi}
\end{equation}
for curve $LM$ in Fig.2;
\begin{equation}\label{51}
  v=\frac{\sin^{2}\varphi}{1+\sin^{2}\varphi},\text{  }
  \alpha=\frac{14\sin^{2}\varphi+8\cos^{4}\varphi}{1+\sin^{2}\varphi}
\end{equation}
for curve $KN$ in Fig.2. Here $\sin^{2}\varphi=\frac{1}{2}(1-c_{+})=\frac{3M_{9}-M_{8}}{6M_{9}-M_{8}-M_{15}}$. Notice that criterion IV is derived along with the curve $A'B'$ in Fig.1. In curve $A'B'$, we have $\sin^{2}\varphi\in[0,\frac{1}{2}]$. The maximal value $\sin^{2}\varphi=\frac{1}{2}$ gives rise to the end points $L,N$ of the curves. The minimal value $\sin^{2}\varphi=0$ gives rise to the end points $K,M$ of the curves.

For the situation of $R_{7}<0$ (namely $\alpha<8u$) described by $STUVS$ in Fig.2, we should have
\begin{equation}\label{52}
   \rho=(1+R'_{1}+R_{7})\varrho_{5\mp}+\frac{-R'_{1}-R_{7}}{2}((|0\rangle\langle0|)^{\otimes4}+(|1\rangle\langle1|)^{\otimes4}),
 \end{equation}
Hence $R_{8}=\mp K\cos^{4}\varphi, R_{15}=\pm K\cos^{2}\varphi\sin^{2}\varphi$ with
$K=\frac{1+6R_{1}+R_{7}}{1+\sin^{2}\varphi}$. With equations (\ref{32a})(\ref{32b}), we have the solution pair $\alpha, v$ as functions of $K, \varphi$, hence
\begin{eqnarray}
  \alpha &=& \frac{7u[1+\sin^2\varphi\mp\cos^2\varphi(8\sin^2\varphi-1)]}{(1+\sin^2\varphi)u\mp\cos^2\varphi(8\sin^2\varphi-1)}, \label{53}\\
  \alpha &=& 8-2v+8(2v-1)\sin^2\varphi.\label{54}
\end{eqnarray}
Where we have used the fact that $1+6R_{1}+R_{7}=1+7R_{7}=8(1-\frac{7u}{\alpha})$. Instead of expressing $v$ as a function of $K, \varphi$, we use $R_{8}/R_{15}$ to obtain equation (\ref{54}).  For the case of $p_{16}=0$, we have $u=1$. So that $\alpha\neq 7$ can only occur when $1+\sin^2\varphi\mp\cos^2\varphi(8\sin^2\varphi-1)=0$. The solutions are $\sin^2\varphi=0,\frac{1}{2}$, respectively. The equations of line sections $GJ$ and $HJ$ in Fig.2 are obtained from (\ref{54}) with $\sin^2\varphi=0,\frac{1}{2}$. Hence criterion IV is also sufficient for states on $GJ$ and $HJ$.

For the states in between $\alpha=8u$ and the physical limitation $p_{1}=0 (\alpha=\frac{14u}{1+u})$, the corresponding $\sin^2\varphi=\frac{1}{8}(4u+1-\sqrt{(4u+21)(4u-3)})$,$\frac{1}{2}$,$0$,$\frac{1}{2}(2-u-\sqrt{(1-u)(7-u)})$ can be derived for $v_{1},v_{2},v_{3},v_{4}$. Thus criterion IV is also sufficient condition for the full separability of the states in the curves.

Hence, for the four qubit highly symmetrical GHZ diagonal state set $\mathcal{S}_{2}$, the necessary and sufficient criterion set is $\mathcal{C}_{2}$=\{Criterion I, Criterion IV\}.

\section{Application to four qubit symmetric GHZ diagonal states}
A GHZ diagonal state is called a symmetric GHZ diagonal state if it is invariant under any qubit permutation. From (\ref{11}), we have $R_{1}=R_{2}=R_{3}=R_{4}=R_{5}=R_{6}$, $R_{9}=R_{10}=R_{11}=R_{12}=R_{13}=R_{14}$ for a four qubit symmetric GHZ diagonal state. Thus the state is specified by $R_{1},R_{7},R_{8},R_{9},R_{15}$. Alternatively, we may use density matrix elements $\rho_{1,1},\rho_{1,16},\rho_{2,2},\rho_{2,15},\rho_{4,4},\rho_{4,13}$ to characterize the state with $\rho_{i,i}=\rho_{17-i,17-i}, \rho_{i,17-i}=\rho_{17-i,i}$ and
\begin{eqnarray}
&\rho_{1,1}=\frac{1}{16}(1+6R_{1}+R_{7}),\text{ }\rho_{1,16}=\frac{1}{16}(R_{8}-6R_{9}+R_{15}),\nonumber\\
&\rho_{2,2}=\rho_{3,3}=\rho_{5,5}=\rho_{8,8}=\frac{1}{16}(1-R_{7}),\nonumber \\
&\rho_{2,15}=\rho_{3,14}=\rho_{5,12}=\rho_{8,9}=\frac{1}{16}(R_{8}-R_{15}),\nonumber \\
&\rho_{4,4}=\rho_{6,6}=\rho_{7,7}=\frac{1}{16}(1-2R_{1}+R_{7}),\nonumber\\
&\rho_{4,13}=\rho_{6,11}=\rho_{7,10}=\frac{1}{16}(R_{8}+2R_{9}+R_{15}).\nonumber
\end{eqnarray}
The normalization condition is $\rho_{1,1}+4\rho_{2,2}+3\rho_{4,4}=\frac{1}{2}$.

Due to criterion III and criterion IV, the equations of the curved surfaces detecting entangled states (outside the surfaces, the states are entangled) are
\begin{equation}\label{62}
\frac{|\rho_{2,15}|}{\Omega}=\frac{1}{2}(1-\frac{\rho_{4,13}}{\Omega}+\sqrt{(1+\frac{\rho_{4,13}}{\Omega})(1+\frac{\rho_{1,16}}{\Omega})})
\end{equation}
for $\rho_{1,16}\geq\rho_{4,13}$ and
\begin{equation}\label{63}
   \frac{|\rho_{2,15}|}{\Omega}=\frac{1}{2}(1+\frac{\rho_{4,13}}{\Omega}+\sqrt{(1-\frac{\rho_{4,13}}{\Omega})(1-\frac{\rho_{1,16}}{\Omega})})
\end{equation}
for $\rho_{1,16}<\rho_{4,13}$, respectively. Where $\Omega$ defined previously can be explained as the minimal diagonal element. We show the surfaces in Fig.3 with $\Omega=\frac{1}{16}$.

If equations (\ref{62}) and (\ref{63}) can be achieved by fully separable states, then the sufficiencies of criterion III and criterion IV are proven for the four qubit symmetric GHZ diagonal states. Without loss of generality, we set $\Omega=\frac{1}{16}$. We may construct the fully separable state from (\ref{48a}). Let
\begin{equation}\label{64}
  \rho=\frac{1}{1+\mu}[\varrho_{+}(\varphi)+\mu\varrho_{-}(\frac{\pi}{2})].
\end{equation}
with $\mu\geq0$ and $\cos(2\varphi)\geq0$. Then $\rho_{1,16}=\frac{1}{16(1+\mu)}[\cos(4\varphi)-\mu]$, $\rho_{2,15}=\frac{1}{16(1+\mu)}[\cos(2\varphi)+\mu]$,
$\rho_{4,13}=\frac{1}{16(1+\mu)}(1-\mu)$. It can be easily checked that equation (\ref{62}) satisfies. Alternatively, if we choose
\begin{equation}\label{65}
  \rho=\frac{1}{1+\mu}[\varrho_{+}(\varphi)+\mu\varrho_{-}(0)],
\end{equation}
with $\cos(2\varphi)\leq0$. We have $\rho_{1,16}=\frac{1}{16(1+\mu)}[\cos(4\varphi)-\mu]$, $\rho_{2,15}=\frac{1}{16(1+\mu)}[\cos(2\varphi)-\mu]$,
$\rho_{4,13}=\frac{1}{16(1+\mu)}(1-\mu)$. Then equation (\ref{63}) satisfies. Similarly, equation (\ref{62})(or (\ref{63})) satisfies by setting $\rho$ to be the mixture of $\varrho_{-}(\varphi)$ and $\varrho_{+}(\frac{\pi}{2})$ (or $\varrho_{+}(0)$ ). Hence, Criterion III and Criterion IV are necessary and sufficient for the full separability of symmetric GHZ diagonal states.

The states in the plane surfaces of Fig.3 can be composed with fully separable states in the following way. Criterion I gives rise to the equation (\ref{45}). It follows the elements of density matrix:
\begin{equation}\label{66}
  \rho_{1,16}=\Omega;\rho_{4,13}=(1-2q_{1})\Omega;\rho_{2,15}=(1-q_{1}-2q_{2})\Omega,
\end{equation}
with $\Omega=\frac{u}{16}$. The probability distribution ($q_{1},q_{2},1-q_{1}-q_{2}$) limits us with $0\leq q_{2}\leq 1-q_{1}$, thus in the plane $\rho_{1,16}=\Omega$ we have two straight line sections $\rho_{2,15}=\frac{1}{2}(\Omega+\rho_{4,13})$ when $q_{2}$ is set to $0$ (its lower bound) and $\rho_{2,15}=-\frac{1}{2}(\Omega+\rho_{4,13})$ when $q_{2}$ is set to $1-q_{1}$ (its upper bound). Hence all the states enclosed by the line sections in the plane are separable sufficiently. More precisely, the states in the triangle with vertices $(\Omega,-\Omega,0)$, $(\Omega,\Omega,\Omega)$, $(\Omega,\Omega,-\Omega)$ in the three-dimensional coordinate system $(\rho_{1,16},\rho_{4,13},\rho_{2,15})$ are separable sufficiently. A state inside the triangle should meet the requirements of $\rho_{2,15}\leq\frac{1}{2}(\Omega+\rho_{4,13})$ and $\rho_{2,15}\geq-\frac{1}{2}(\Omega+\rho_{4,13})$, they are just the conditions $\frac{R_{8}}{R_{9}}\leq 1, \frac{R_{15}}{R_{9}}\leq 1$ in (\ref{23}) for Criterion I.  Similarly, the states in the triangle with vertices $(-\Omega,\Omega,0)$, $(-\Omega,-\Omega,-\Omega)$, $(-\Omega,-\Omega,\Omega)$ in the coordinate system are separable sufficiently. Hence, Criterion I is necessary and sufficient for the full separability of symmetric GHZ diagonal states.

The intersection of the curved surface (\ref{62}) and the plane $\rho_{4,13}=\Omega$ is a parabola
 \begin{equation}\label{67}
  \frac{|\rho_{2,15}|}{\Omega}=\sqrt{\frac{1}{2}(1+\frac{\rho_{1,16}}{\Omega})}
 \end{equation}
 with $\Omega\geq\rho_{1,16}\geq-\Omega$.
 The states enclosed by the parabola should meet the requirement of $\frac{|\rho_{2,15}|}{\Omega}\leq\sqrt{\frac{1}{2}(1+\frac{\rho_{1,16}}{\Omega})}$, it is just the condition $\frac{R_{8}R_{15}}{R_{9}^2}\geq 1$ in (\ref{23}) for Criterion II. In the plane of $\rho_{3,14}=\Omega$, the states enclosed by parabola (\ref{67}) can be decomposed as probability mixture of the states on the parabola, thus they are fully separable due to the full separability of the states on the parabola. Hence Criterion II is necessary and sufficient for the states enclosed by parabola (\ref{67}) on the plane $\rho_{3,14}=\Omega$. Similarly we can show that Criterion II is necessary and sufficient for the states enclosed by parabola $\frac{|\rho_{2,15}|}{\Omega}=\sqrt{\frac{1}{2}(1-\frac{\rho_{1,16}}{\Omega})}$ on the plane $\rho_{3,14}=-\Omega$.

All the boundaries determined by the necessary criteria for the four qubit symmetric GHZ diagonal states are proved to be also sufficient, since the states on boundaries are proved to be fully separable. Outside the boundaries, the states are entangled.
\begin{figure}[tpb]
\includegraphics[ trim=0.000000in 0.000000in -0.138042in 0.000000in,
height=2.5in, width=3.5in]{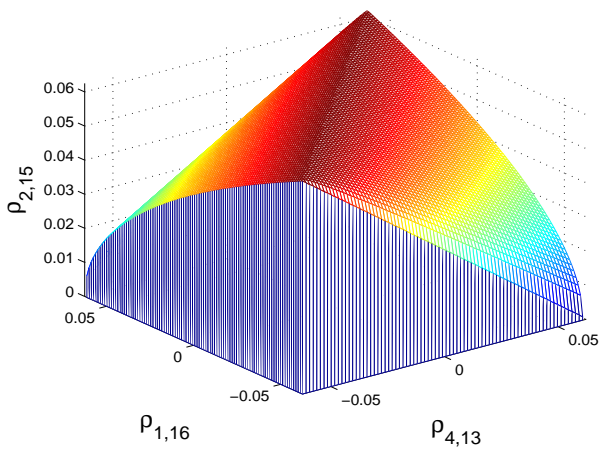}
\caption{(Color on line)The curved surfaces and the other plane surfaces are the boundaries of fully separable state set. We show the positive $\rho_{2,15}$ in the figure with $\Omega=\frac{1}{16}$. The mirror reflection of the figure with respect to $\rho_{1,16}-\rho_{4,13}$ plane is not shown.}
\end{figure}

Hence, for the four qubit symmetrical GHZ diagonal state set $\mathcal{S}_{3}$, the necessary and sufficient criterion set is $\mathcal{C}_{3}$=\{Criterion I,Criterion II,Criterion III,Criterion IV\}.
\section{Discussion}
 The PPT criterion of separability for the four qubit GHZ diagonal states is
 \begin{equation}\label{67a}
   \max_{i}|\rho_{i,17-i}|\leq \min_{j}\rho_{j,j},
 \end{equation}
 the maximal absolute anti-diagonal element does not exceed the minimal diagonal element. This criterion can also be obtained with proper $M_{i}$ subjecting to $|M_{i}|=1$. More explicitly, from (\ref{3}) we have
 \begin{equation}\label{68}
 g(\boldsymbol{\varphi})=\cos(\varphi_{1}+\varphi_{2}+\varphi_{3}+\varphi_{4})\leq 1,
\end{equation}
by setting $M_{8}=M_{15}=1, M_{9}=...=M_{14}=-1$. Thus $\tilde{g}=1$. Notice that $\sum_{i=8}^{15}M_{i}R_{i}=16\rho_{1,16}$. Changing the sign of $\varphi_{j}$ (j=1,2,3,4) leads to different assignment of signs for $M_{i}$ (i=8,...,15) while keeping $\tilde{g}=1$. The maximal $\sum_{i=8}^{15}M_{i}R_{i}$ is $16\max|\rho_{i,17-i}|$ (i=1,...16).

For the four qubit symmetric GHZ diagonal states, the PPT criterion (\ref{67a}) defines a cube. The surface of cube coincides with the some parts of border surface between separable and entangled states.

For GHZ diagonal states, there is a sufficient condition for the coincidence of the PPT criterion and the full separability criterion\cite{Kay}. The condition can be written as
\begin{equation}\label{69}
  R_{9}R_{15}\leq 0 \text{  and } R_{9}R_{8}\leq 0.
\end{equation}
  The border states in the planes $\rho_{4,13}=-\Omega$ and $\rho_{4,13}=\Omega$ do not satisfy the condition (\ref{69}). On plane surfaces $\rho_{1,16}=\pm\Omega$, the border states form two triangles. Only some of these states satisfy the condition (\ref{69}), they form two less size triangles.

\section{Conclusion}

We have found that the set of necessary and sufficient criteria of separability has a subset structure. The smallest criterion set $\mathcal{C}_{1}$ has one criterion (criterion I), it can detect the full separability of the generalized Werner state (GHZ state mixed with white noise or identity) set $\mathcal{S}_{1}$ necessarily and sufficiently. The state set $\mathcal{S}_{2}$ is the set of four qubit highly symmetric GHZ diagonal states. It includes $\mathcal{S}_{1}$ as its subset. We have found the criterion set $\mathcal{C}_{2}$ with two criteria (criterion I and Criterion IV) which can detect the full separability of states in $\mathcal{S}_{2}$. The criterion set $\mathcal{C}_{1}$ is a subset of $\mathcal{C}_{2}$. Criterion set $\mathcal{C}_{3}$ has four criteria: Criterion I,Criterion II, Criterion III, Criterion IV. Criterion set $\mathcal{C}_{3}$ is necessary and sufficient for the state set $\mathcal{S}_{3}$ of four qubit symmetric GHZ diagonal states. We have necessary and sufficient criterion set $\mathcal{C}_{i}$ for state set $\mathcal{S}_{i}$ for $i=1,2,3$, with $\mathcal{C}_{2}$ being the subset of $\mathcal{C}_{3}$ and the superset of $\mathcal{C}_{1}$, $\mathcal{S}_{2}$ being the subset of $\mathcal{S}_{3}$ and the superset of $\mathcal{S}_{1}$. The nest structure of the criterion set suggests us a way of developing criterion set for large state set from some seed state set.

We have used the two step of optimizations to analytically obtain $\mathcal{C}_{3}$ when the state is a four qubit symmetric GHZ diagonal state. We proved the sufficiency of $\mathcal{C}_{3}$ for these symmetric states by explicitly constructing these states with fully separable states. All boundaries of the fully separable state set of four qubit symmetric GHZ diagonal states have been found by the criteria. We also have studied the sufficient condition for the coincidence of PPT criterion and fully separable criterion.

\section*{Acknowledgement}

Supported by the National Natural Science Foundation of China ( Grant Nos: 61871347 and 11375152) are gratefully acknowledged.

\end{document}